\documentclass[%
 aip,
 amsmath,amssymb,
 reprint,%
]{revtex4-1}
\usepackage{dcolumn}
\usepackage{bm}
\usepackage[utf8]{inputenc}
\usepackage[T1]{fontenc}
\usepackage{mathptmx}
\usepackage[normalem]{ulem}
\usepackage{mathtools}
\usepackage{graphicx}
\usepackage{epstopdf}
\usepackage{lmodern}
\usepackage{xfrac}
\usepackage[breaklinks]{hyperref}
\usepackage{mathtools}
\usepackage{xcolor}

\begin{document}

\title{Symmetrical clock synchronization with time-correlated photon pairs}

\author{Jianwei~Lee}
\affiliation{Centre for Quantum Technologies, National University of Singapore, 3 Science Drive 2, Singapore 117543}

\author{Lijiong~Shen}
\affiliation{Centre for Quantum Technologies, National University of Singapore, 3 Science Drive 2, Singapore 117543}
\affiliation{Department of Physics, National University of Singapore, 2 Science Drive 3, Singapore 117551}

\author{Alessandro~Cer\`{e}}
\affiliation{Centre for Quantum Technologies, National University of Singapore, 3 Science Drive 2, Singapore 117543}

\author{James~Troupe}
\affiliation{Applied Research Laboratories, The University of Texas at Austin, Austin, Texas}

\author{Antia~Lamas-Linares}
\affiliation{Texas Advanced Computing Center, The University of Texas at Austin, Austin, Texas}
\affiliation{Centre for Quantum Technologies, National University of Singapore, 3 Science Drive 2, Singapore 117543}

\author{Christian~Kurtsiefer}
\affiliation{Centre for Quantum Technologies, National University of Singapore, 3 Science Drive 2, Singapore 117543}
\affiliation{Department of Physics, National University of Singapore, 2 Science Drive 3, Singapore 117551}

\email[]{christian.kurtsiefer@gmail.com}
\date{\today}
\begin{abstract} 
We demonstrate a point-to-point clock synchronization protocol based on bidirectionally exchanging photons produced in spontaneous parametric down conversion (SPDC). The technique exploits tight timing correlations between photon pairs to achieve a precision of 51\,ps in 100\,s with count rates of order 200\,s$^{-1}$. The protocol is distance independent, secure against symmetric delay attacks and provides a natural complement to techniques based on Global Navigation Satellite Systems (GNSS). The protocol works with mobile parties and can be augmented to provide authentication of the timing signal via a Bell inequality check.
\end{abstract}


\maketitle

\section{Introduction}
The ability to synchronize remote clocks plays an important role in our
infrastructure, from maintaining coherence in the electrical grid, to allowing
precise positioning and navigation, high speed trading, and distributed data processing.
However, many of the techniques to establish and maintain this time synchronization have been shown to be susceptible to interference by malicious parties~\cite{shepard:12,Bhatti:2017}, which can for example spoof the legitimate timing signal introducing unaccounted for delays, and thus introducing an error in the calculated time difference between the remote clocks.

In most protocols, remote parties deduce their clock offset 
by measuring signal propagation times with their devices
and comparing the result with a trusted value~\cite{mills:1991,PTP,Moreira:2009}. 
Protocol security then relies on an independent characterization of
propagation times~\cite{narula:17}, which can be difficult for mobile parties or under changing conditions.

In this work, we describe a distance-independent protocol using counter-propagating single photons originating from SPDC pairs.
Tight time correlations of photon pairs generated from spontaneous parametric down-conversion (SPDC) enable precise synchronization.
Similar to existing bidirectional protocols,
clock offsets are determined independently of signal propagation times for a symmetrical communication channel~\cite{narula:17}.
The single-photon regime allows, in principle, an additional security layer 
by testing a Bell inequality with entangled photons to verify the origin of the timing signal.

While clock synchronization based on SPDC has been demonstrated, previous works require knowing \textit{a priori} the signal propagation times~\cite{valencia:04,Marcikic:2006,ho:09}, 
or controlling them with a balanced interferometer~\cite{quan:16}.
In contrast, we use a simple point-to-point single mode optical connection, which is typical of quantum communication scenarios
As a proof-of-principle demonstration, we synchronize two remote rubidium clocks while changing their relative separation, simulating a symmetric channel delay attack.

\section{Time synchronization protocol}
\begin{figure}
  \centering
  \includegraphics[width=\columnwidth]{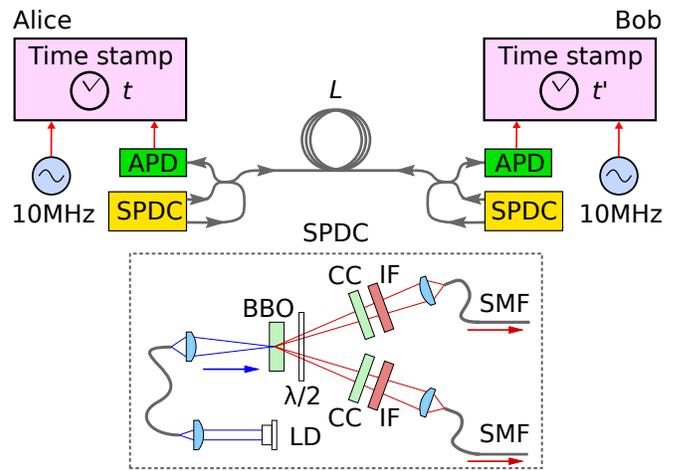}
  \caption{\label{fig:setup}
  Clock synchronization setup. Alice and Bob each have a source of time-correlated photon pairs based on spontaneous parametric down-conversion (SPDC), and an avalanche photodetector (APD). One photon of the pair is detected locally, 
  while the other
  photon is sent through a single mode fiber of length~$L$ to be detected on the remote side. Times of arrival for all detected photons are recorded at each side with respect to the local clock, each locked to a rubidium frequency reference.
  The inset shows the optical setup of a SPDC source~\cite{Kwiat:1995ub}. LD: laser diode, BBO: $\beta$-Barium~Borate, CC: compensation crystals, SMF: single mode fiber, $\lambda$/2: half-wave plate. 
  }
\end{figure}
The protocol involves two parties, Alice and Bob, connected by a single mode optical channel. 
Each party has an SPDC source producing photons pairs, one photon is detected locally, while the other is sent and detected on the remote side~(see Fig.~\ref{fig:setup}). Every photodetection event is time tagged according to a local clock which assigns time stamps $t$ and $t'$. 

For a propagation time $\Delta t_{AB}$ from Alice to Bob, and $\Delta t_{BA}$ in the other direction, the detection time differences are
\begin{equation}\label{eq:deltaone}
    t'-t = \Delta t_{AB} + \delta \quad\mathrm{and}\quad
    t - t' = \Delta t_{BA} - \delta\,
\end{equation}
for the photon pairs originating from Alice and Bob, respectively.
The sequence of photodetection events on each side are described by
\begin{equation}
a(t)=\sum_i \delta\left (t-t_i\right) \quad\mathrm{and}\quad
b(t')=\sum_j \delta\left (t'-t'_j\right)\,.
\end{equation}
Due to tight time correlations present during pair generation, the cross correlation
\begin{equation}\label{eq:C_AB}
c_{AB}(\tau)=(a\star b)(\tau)=\int a(t)b(t+\tau)\mathrm{d}t
\end{equation}
will show two peaks at
\begin{equation}
  \tau_{AB} = \delta + \Delta t_{AB} \quad\mathrm{and}\quad
  \tau_{BA} = \delta - \Delta t_{BA}\,
\end{equation}
for the pairs created by Alice and Bob.
A round-trip time $\Delta T$ for photons can be calculated using the inter-peak separation,
\begin{equation}\label{eq:round_trip}
\Delta T = \Delta t_{AB} + \Delta t_{BA} = \tau_{AB}-\tau_{BA}. 
\end{equation}
If the propagation times in the two directions are the same, $\Delta
t_{AB}=\Delta t_{BA}$, they do not contribute to the clock offset
\begin{equation}\label{eq:offset}
\delta = \frac{1}{2}\left (\tau_{AB}+\tau_{BA}\right )\,,
\end{equation}
which is calculated directly from the midpoint of the two peaks.
In this way, the protocol is inherently robust against symmetric changes in channel propagation times.

As is the norm in quantum key distribution (QKD)~\cite{scarani:09}, the time stamps are transmitted through a classical public authenticated channel, while the quantum channel is supposed to be under the control of a malicious adversary.

\section{Experiment}\label{sec:experiment}
Time-correlated photon pairs are generated by two identical SPDC sources (Fig.~\ref{fig:setup}).
The output of a laser diode (power $\approx$10\,mW, central wavelength 405\,nm) is coupled into a single mode optical fiber for spatial mode filtering and focused to a beam waist of 80\,$\mu$m into a 2\,mm thick $\beta$-Barium~Borate crystal cut for non-collinear type-II phase matching~\cite{Kwiat:1995ub}.
Down-converted photons at 810\,nm are coupled into two single
mode fibers; with an overall detected pair rate of about~$200$\,s$^{-1}$.

Fiber beam splitters separate the photon pairs so that one photon is detected locally with an avalanche photodetector (APD), while the other photon is transmitted to the remote party. 
Time-stamping units with nominal resolution~$\approx4$\,ps assign detection times
 $t$ and $t'$ to the events detected at Alice and Bob, respectively.
 
To resolve the coincidence peaks (FWHM $\approx$ 500\,ps), we obtain
$c_{AB}(\tau = t' - t)$ with coarse ($\approx 2\,\mu$s) and fine ($\approx
16\,$ps) resolutions separately~\cite{Marcikic:2006}. 

To extract the peak positions $\tau_{AB}$ and $\tau_{BA}$,
we fit $c_{AB}(\tau)$ to a linear combination of two peak profiles $V(\tau)$,
\begin{equation}\label{eq:fitmodel}
  c_{AB}(\tau) = a_0 + a_1\,V(\tau-\tau_{AB}) + a_2\,V(\tau-\tau_{BA})\,,
\end{equation}
where $a_0$ denotes background coincidences, $a_{1,2}$ detected pairs,
and $V(\tau)$ is a pseudo-Voigt distribution~\cite{Wertheim:1974}
\begin{equation}\label{eq:pseudovoigt}
  V(\tau) = 
  (1-f)\,G\left(\tau,\frac{\sigma}{\sqrt{2\ln2}}\right) + f\,L\left(\tau,\sigma\right)\,.
\end{equation}
Values of $f = 0.2$ and $\sigma = 290$\,ps best characterize the timing jitter 
of the combined photodetection and time-stamping system, and 
$\tau_{AB},\tau_{BA}$ from the fit fix $\delta$ and $\Delta T$ through equations~\ref{eq:round_trip} and~\ref{eq:offset}.

\section{Results}

\subsection{Synchronization precision}\label{sec:precision}
\begin{figure}
  \centering
  \includegraphics[width=\columnwidth]{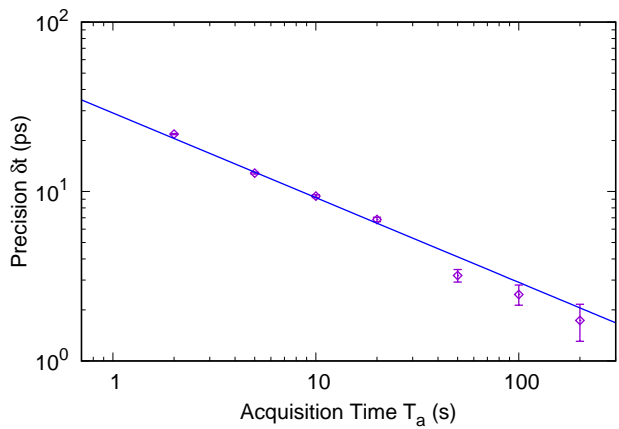}
  \caption{\label{fig:precision} 
  Standard deviation (precision $\delta t$) of the measured offset between two clocks. 
  Both clocks are locked to the same frequency reference.
  Solid line: Least-squares fit to a model where $\delta t$ follows Poisson statistics and improves with acquisition time $T_a$.
  Error bars: precision uncertainty due to errors from fitting $c_{AB}$ to our model in Eq.~\ref{eq:fitmodel}.   
  }
\end{figure}
To demonstrate the independence of the protocol from the clock separation,
we first determine the minimum resolvable separation ($v\,\delta t / 2$),
where $v$ is the propagation speed of light in the fiber, and $\delta t$ is the precision (1 standard deviation) of measuring a fixed offset.

To characterize the precision $\delta t$, 
we accumulate offset measurements between two clocks locked to a common frequency reference (Stanford Research Systems FS725), separated by a constant fiber length $L = 1.7$\,m.
The standard deviation of the measured offset depends on the detector timing response $V(\tau=0)$, pair rate $R$ and acquisition time $T_a$ according to~\cite{Rider:1960}
\begin{equation}\label{eq:PoissonianStats}
  \delta t = \frac{1}{\sqrt{2}} \, \frac{1}{2\,V(\tau=0)}\,\frac{1}{\sqrt{R\,T_a}}.
\end{equation}
Figure~\ref{fig:precision} shows the precision of the measured offset for various $T_a$, extracted from time stamps recorded over 1\,hour.
The timing response $V(\tau=0)=1.65(5)$~ns$^{-1}$, extracted from a fit to the data using Eq.~\ref{eq:PoissonianStats}, closely describe our detectors ($1.5$~ns$^{-1}$). 

For an acquisition lasting several seconds, a precision of a few picoseconds
limits the minimum resolvable clock separation to the millimeter scale.  
To demonstrate that the protocol is secure against symmetric channel delay
attacks, we change the propagation length over several meters during
synchronization --- three orders of magnitude larger than the minimum resolvable length-scale.

\subsection{Distance-independent clock synchronization}\label{sec:commonfreq}
\begin{figure}
  \centering
  \includegraphics[width=\columnwidth]{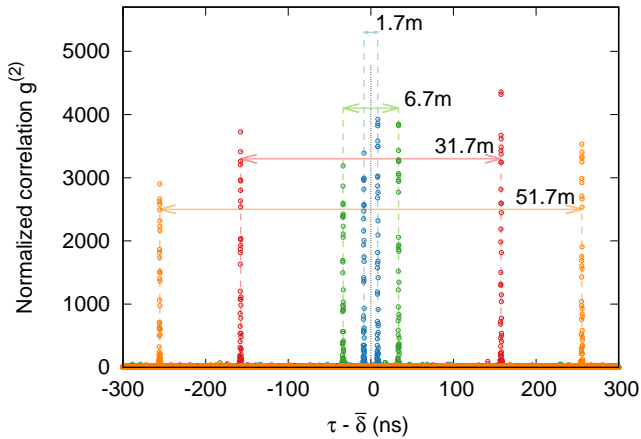}
  \caption{\label{fig:g2}
  Timing correlations of Alice and Bob's detection events normalized to background coincidences.
  During the measurement, four fibers of lengths $L$ were used to change the separation between Alice and Bob.
  For every~$L$, the correlation measurement yields two coincidence peaks,
  one for each source.
  The time separation between peaks corresponds to the round-trip time~$\Delta T$, and the midpoint is the offset between the clocks~$\delta$.
  The time axis is shifted by $\overline{\delta}$,
  the average value of the four~$\delta$ calculated for four different~$L$.
  }
\end{figure}

\begin{figure}
  \centering
  \includegraphics[width=\columnwidth]{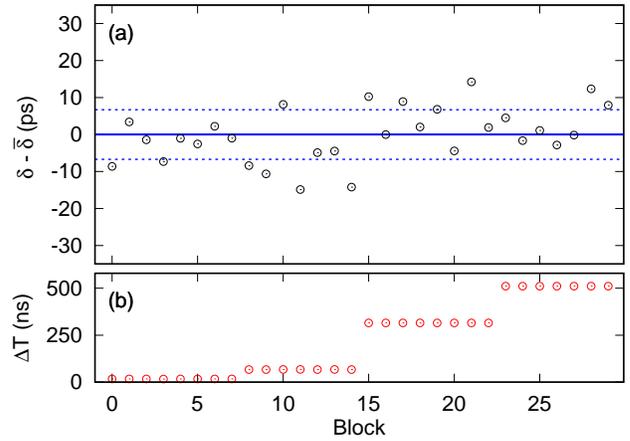}
  \caption{\label{fig:offsetvsdist}
  (a) Measured offset $\delta$ between two clocks, both locked on the same frequency reference. 
  Each value of $\delta$ was evaluated from measuring photon pair timing correlations from a block of photodetection times recorded by Alice and Bob. Each block is 20\,s long.
  The continuous line indicates the average offset $\overline{\delta}$.
  Dashed lines: one standard deviation. 
  (b) The round-trip time $\Delta T$ was changed using different fiber lengths.
  }
\end{figure}

To simulate a symmetric channel delay attack, we
impose different propagation distances using different fibers of length $L=1.7$\,m, $6.7$\,m, 31.7\,m, and~51.7\,m.
Figure~\ref{fig:g2} shows $g^{(2)}(\tau)$, the cross-correlation $c_{AB}(\tau)$ normalized to background coincidences, acquired from the time stamps recorded over 20~mins. 
To detect changes in the clock offset throughout the acquisition, we split the
time-stamped events into blocks of 20\,s.
Figure~\ref{fig:offsetvsdist} shows the clock offset~$\delta$ and round-trip time~$\Delta T$ for every block. 
Throughout the acquisition, the offset was measured to within $7$\,ps, comparable to the precision obtained with a constant round-trip time~(Fig.~\ref{fig:precision}).
With no significant correlation between the measured clock offset and the propagation distance ($\lesssim 0.12$\,ps\,m$^{-1}$), we conclude that for measuring a fixed offset, the protocol is robust against symmetric channel delay attacks.

\subsection{Distance-independent clock synchronization with independent clocks}\label{sec:difffreq}
\begin{figure}
  \centering
  \includegraphics[width=\columnwidth]{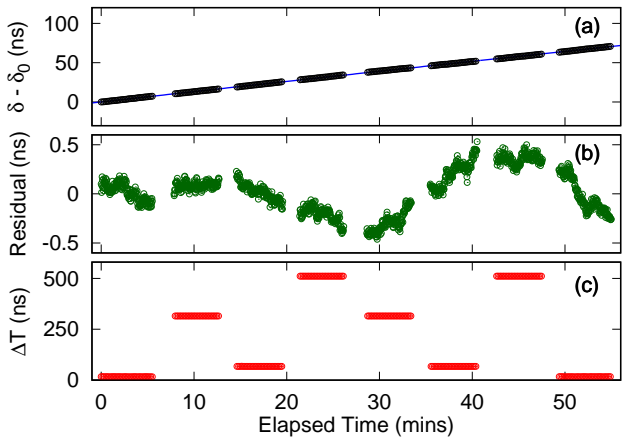}
  \caption{\label{fig:offsetvsdistsepclocks}
  (a) Measured offset $\delta$ between two clocks with different frequency references. 
  Each value of $\delta$ was evaluated from measuring photon pair timing correlations for 2\,s.
  The offset measured at the beginning is $\delta_0$. 
  Continuous blue line: fit used to extract the relative frequency accuracy ($\approx4\times10^{-11}$) between the clocks.
  (b) Residual of the fit fluctuates due to the intrinsic instability of the individual frequency references. 
  (c) The round-trip time $\Delta T$ was changed using four different fiber lengths.
  }
\end{figure}
To examine a more realistic scenario, we provide each time-stamping unit with an independent frequency reference (both Stanford Research Systems FS725), 
resulting in a clock offset that drifts with time~$\delta\rightarrow\delta(t)$. 

The frequency references have a nominal relative frequency accuracy $d_0 < 5\times10^{-11}$.
We evaluate the offset from the time stamps every $T_a = 2$\,s so that the drift ($\approx100$\,ps) is much smaller than the FWHM of each coincidence peak.
This allows extracting the peak positions from $c_{AB}$ with the model in Eq.~\ref{eq:fitmodel}.

We again simulate a symmetric channel delay attack by changing $L$ every 5\,mins.
Figure~\ref{fig:offsetvsdistsepclocks} shows the measured $\delta(t)$ which appears to follow a continuous trend over different round-trip times, indicating that the delay attacks were ineffective. 
Discontinuities in $\delta(t)$ correspond to periods when fibers were changed. 
 
To verify that meaningful clock parameters can be extracted from $\delta(t)$ despite the attack, we fit the data to a parabola $a\,t^2 + d\,t + b$, where $a$, $d$ and $b$ represent the relative aging, frequency accuracy and bias of the frequency references, respectively~\cite{Xu:2016ji}. 
The resulting relative frequency accuracy between the clocks, $d = 4.05(7)\times10^{-11}$, agrees with the nominal relative frequency accuracy $d_0$ of our frequency references.
The residual of the fit, $r(t)$, 
fluctuates~(Allan deviation~$=1.1\times10^{-12}$, time deviation TDEV = 45\,ps, in 100\,s) mainly due to the intrinsic instabilities of our frequency references~($<2\times10^{-12}$).
Negligible correlation between $r(t)$ and propagation distance ($\lesssim 0.78$\,ps\,m$^{-1}$) demonstrates the distance-independence of this protocol.

The standard deviation ($\delta t\approx51$\,ps) of the fast fluctuating
component of $r(t)$ suggests that  
the clocks can be synchronized to a precision 
comparable to the time deviation 
of our frequency references in 100\,s.
This integration time reduces with detectors with a lower timing jitter,
higher efficiency, a higher path transmission, and with brighter pair
sources~(Eq.~\ref{eq:PoissonianStats}).

\section{Protocol Security}\label{sec:security}
Although not demonstrated in this work, Alice and Bob can verify the origin of each photon by synchronizing with polarization-entangled photon pairs and performing a Bell measurement to check for correspondence between the local and transmitted photons. 
As is the case in QKD scenarios~\cite{ekert:92}, if the signal is copied (cloned) or the entangled degree of freedom is otherwise disturbed, the extent of the interference can be bounded via a Bell inequality.
For this measurement, the setup in Fig.~\ref{fig:setup} should be modified such that 
the detectors are preceded by a polarization measurement in the appropriate basis and that measurement result is added to the time stamp information transmitted through the classical channel.
This modification addresses the issue of spoofing in current classical synchronization protocols.

In addition, we made the strong assumption that the photon propagation times in both directions were equal ($\Delta t_{AB} = \Delta t_{BA}$).
Without this assumption, the offset derived from Eq.~\ref{eq:offset} becomes
\begin{equation}
  \delta = \frac{1}{2}\,[(\tau_{AB} + \tau_{BA}) - (\Delta t_{AB} - \Delta t_{BA})].
\end{equation} 
Therefore, the offset can no longer be obtained from the midpoint between $\tau_{AB}$ and $\tau_{BA}$. 

We note that while creating an asymmetric channel for a classical signal is routine given the ability to split and amplify the signal at will; in the case of entangled photons produced at random times, making an asymmetric channel implies breaking the reciprocity of the channel. This is possible, via for example magneto-optical effects such as found in optical circulators. Detecting this attack is the subject of ongoing research~\cite{Troupe:2018}.

\section{Conclusion}
We have demonstrated a protocol for synchronizing two remote clocks with time-correlated photon pairs generated from SPDC.
By assuming symmetry in the synchronization channel,
our protocol does not require \textit{a priori} knowledge of the relative distance or propagation times between two parties, providing security against symmetric channel delay attacks and timing signal authentication via the measurement of a Bell inequality.

We observe a synchronization precision of 51\,ps within 100\,s, 
which is comparable to the time deviation arising from the intrinsic instability of our frequency references, even with relatively low pair rates ($\approx 200$~s$^{-1}$). 

The protocol lends itself particularly well to synchronization tasks performed between mobile stations (e.g., between satellites and ground stations) where photon rates are typically low, and propagation times are constantly changing. 
Since the protocol is based on existing quantum communication techniques, 
it provides a natural complement to Global Navigation Satellite Systems
(GNSS) and would be a natural fit to future quantum networks with the ability to distribute entanglement.

\bibliographystyle{apsrev4-1}

\begin{thebibliography}{17}%
\makeatletter
\providecommand \@ifxundefined [1]{%
 \@ifx{#1\undefined}
}%
\providecommand \@ifnum [1]{%
 \ifnum #1\expandafter \@firstoftwo
 \else \expandafter \@secondoftwo
 \fi
}%
\providecommand \@ifx [1]{%
 \ifx #1\expandafter \@firstoftwo
 \else \expandafter \@secondoftwo
 \fi
}%
\providecommand \natexlab [1]{#1}%
\providecommand \enquote  [1]{``#1''}%
\providecommand \bibnamefont  [1]{#1}%
\providecommand \bibfnamefont [1]{#1}%
\providecommand \citenamefont [1]{#1}%
\providecommand \href@noop [0]{\@secondoftwo}%
\providecommand \href [0]{\begingroup \@sanitize@url \@href}%
\providecommand \@href[1]{\@@startlink{#1}\@@href}%
\providecommand \@@href[1]{\endgroup#1\@@endlink}%
\providecommand \@sanitize@url [0]{\catcode `\\12\catcode `\$12\catcode
  `\&12\catcode `\#12\catcode `\^12\catcode `\_12\catcode `\%12\relax}%
\providecommand \@@startlink[1]{}%
\providecommand \@@endlink[0]{}%
\providecommand \url  [0]{\begingroup\@sanitize@url \@url }%
\providecommand \@url [1]{\endgroup\@href {#1}{\urlprefix }}%
\providecommand \urlprefix  [0]{URL }%
\providecommand \Eprint [0]{\href }%
\providecommand \doibase [0]{http://dx.doi.org/}%
\providecommand \selectlanguage [0]{\@gobble}%
\providecommand \bibinfo  [0]{\@secondoftwo}%
\providecommand \bibfield  [0]{\@secondoftwo}%
\providecommand \translation [1]{[#1]}%
\providecommand \BibitemOpen [0]{}%
\providecommand \bibitemStop [0]{}%
\providecommand \bibitemNoStop [0]{.\EOS\space}%
\providecommand \EOS [0]{\spacefactor3000\relax}%
\providecommand \BibitemShut  [1]{\csname bibitem#1\endcsname}%
\let\auto@bib@innerbib\@empty
\bibitem [{\citenamefont {Shepard}\ \emph {et~al.}(2012)\citenamefont
  {Shepard}, \citenamefont {Humphreys},\ and\ \citenamefont
  {Fansler}}]{shepard:12}%
  \BibitemOpen
  \bibfield  {author} {\bibinfo {author} {\bibfnamefont {D.~P.}\ \bibnamefont
  {Shepard}}, \bibinfo {author} {\bibfnamefont {T.~E.}\ \bibnamefont
  {Humphreys}}, \ and\ \bibinfo {author} {\bibfnamefont {A.}~\bibnamefont
  {Fansler}},\ }\href {\doibase https://doi.org/10.1016/j.ijcip.2012.09.003}
  {\bibfield  {journal} {\bibinfo  {journal} {International Journal of Critical
  Infrastructure Protection}\ }\textbf {\bibinfo {volume} {5}},\ \bibinfo
  {pages} {146} (\bibinfo {year} {2012})}\BibitemShut {NoStop}%
\bibitem [{\citenamefont {Bhatti}\ and\ \citenamefont
  {Humphreys}(2017)}]{Bhatti:2017}%
  \BibitemOpen
  \bibfield  {author} {\bibinfo {author} {\bibfnamefont {J.}~\bibnamefont
  {Bhatti}}\ and\ \bibinfo {author} {\bibfnamefont {T.~E.}\ \bibnamefont
  {Humphreys}},\ }\href {\doibase 10.1002/navi.183} {\bibfield  {journal}
  {\bibinfo  {journal} {Navigation}\ }\textbf {\bibinfo {volume} {64}},\
  \bibinfo {pages} {51} (\bibinfo {year} {2017})},\ \Eprint
  {http://arxiv.org/abs/https://onlinelibrary.wiley.com/doi/pdf/10.1002/navi.183}
  {https://onlinelibrary.wiley.com/doi/pdf/10.1002/navi.183} \BibitemShut
  {NoStop}%
\bibitem [{\citenamefont {Mills}(1991)}]{mills:1991}%
  \BibitemOpen
  \bibfield  {author} {\bibinfo {author} {\bibfnamefont {D.~L.}\ \bibnamefont
  {Mills}},\ }\href {\doibase 10.1109/26.103043} {\bibfield  {journal}
  {\bibinfo  {journal} {IEEE Transactions on Communications}\ }\textbf
  {\bibinfo {volume} {39}},\ \bibinfo {pages} {1482} (\bibinfo {year}
  {1991})}\BibitemShut {NoStop}%
\bibitem [{PTP(2009)}]{PTP}%
  \BibitemOpen
  \href {\doibase 10.1109/IEEESTD.2009.4839002} {\bibfield  {journal} {\bibinfo
   {journal} {IEC 61588:2009(E)}\ ,\ \bibinfo {pages} {C1}} (\bibinfo {year}
  {2009})}\BibitemShut {NoStop}%
\bibitem [{\citenamefont {Moreira}\ \emph {et~al.}(2009)\citenamefont
  {Moreira}, \citenamefont {Serrano}, \citenamefont {Wlostowski}, \citenamefont
  {Loschmidt},\ and\ \citenamefont {Gaderer}}]{Moreira:2009}%
  \BibitemOpen
  \bibfield  {author} {\bibinfo {author} {\bibfnamefont {P.}~\bibnamefont
  {Moreira}}, \bibinfo {author} {\bibfnamefont {J.}~\bibnamefont {Serrano}},
  \bibinfo {author} {\bibfnamefont {T.}~\bibnamefont {Wlostowski}}, \bibinfo
  {author} {\bibfnamefont {P.}~\bibnamefont {Loschmidt}}, \ and\ \bibinfo
  {author} {\bibfnamefont {G.}~\bibnamefont {Gaderer}},\ }in\ \href {\doibase
  10.1109/ISPCS.2009.5340196} {\emph {\bibinfo {booktitle} {2009 International
  Symposium on Precision Clock Synchronization for Measurement, Control and
  Communication}}}\ (\bibinfo {year} {2009})\ pp.\ \bibinfo {pages}
  {1--5}\BibitemShut {NoStop}%
\bibitem [{\citenamefont {Narula}\ and\ \citenamefont
  {Humphreys}(2018)}]{narula:17}%
  \BibitemOpen
  \bibfield  {author} {\bibinfo {author} {\bibfnamefont {L.}~\bibnamefont
  {Narula}}\ and\ \bibinfo {author} {\bibfnamefont {T.~E.}\ \bibnamefont
  {Humphreys}},\ }\href {\doibase 10.1109/JSTSP.2018.2835772} {\bibfield
  {journal} {\bibinfo  {journal} {IEEE Journal of Selected Topics in Signal
  Processing}\ }\textbf {\bibinfo {volume} {12}},\ \bibinfo {pages} {749}
  (\bibinfo {year} {2018})}\BibitemShut {NoStop}%
\bibitem [{\citenamefont {Valencia}\ \emph {et~al.}(2004)\citenamefont
  {Valencia}, \citenamefont {Scarcelli},\ and\ \citenamefont
  {Shih}}]{valencia:04}%
  \BibitemOpen
  \bibfield  {author} {\bibinfo {author} {\bibfnamefont {A.}~\bibnamefont
  {Valencia}}, \bibinfo {author} {\bibfnamefont {G.}~\bibnamefont {Scarcelli}},
  \ and\ \bibinfo {author} {\bibfnamefont {Y.}~\bibnamefont {Shih}},\ }\href
  {https://doi.org/10.1063/1.1797561} {\bibfield  {journal} {\bibinfo
  {journal} {Applied Physics Letters}\ }\textbf {\bibinfo {volume} {85}},\
  \bibinfo {pages} {2655} (\bibinfo {year} {2004})}\BibitemShut {NoStop}%
\bibitem [{\citenamefont {Marcikic}\ \emph {et~al.}(2006)\citenamefont
  {Marcikic}, \citenamefont {Lamas-Linares},\ and\ \citenamefont
  {Kurtsiefer}}]{Marcikic:2006}%
  \BibitemOpen
  \bibfield  {author} {\bibinfo {author} {\bibfnamefont {I.}~\bibnamefont
  {Marcikic}}, \bibinfo {author} {\bibfnamefont {A.}~\bibnamefont
  {Lamas-Linares}}, \ and\ \bibinfo {author} {\bibfnamefont {C.}~\bibnamefont
  {Kurtsiefer}},\ }\href {\doibase 10.1063/1.2348775} {\bibfield  {journal}
  {\bibinfo  {journal} {Applied Physics Letters}\ }\textbf {\bibinfo {volume}
  {89}},\ \bibinfo {pages} {101122} (\bibinfo {year} {2006})},\ \Eprint
  {http://arxiv.org/abs/https://doi.org/10.1063/1.2348775}
  {https://doi.org/10.1063/1.2348775} \BibitemShut {NoStop}%
\bibitem [{\citenamefont {Ho}\ \emph {et~al.}(2009)\citenamefont {Ho},
  \citenamefont {Lamas-Linares},\ and\ \citenamefont {Kurtsiefer}}]{ho:09}%
  \BibitemOpen
  \bibfield  {author} {\bibinfo {author} {\bibfnamefont {C.}~\bibnamefont
  {Ho}}, \bibinfo {author} {\bibfnamefont {A.}~\bibnamefont {Lamas-Linares}}, \
  and\ \bibinfo {author} {\bibfnamefont {C.}~\bibnamefont {Kurtsiefer}},\
  }\href {http://stacks.iop.org/1367-2630/11/i=4/a=045011} {\bibfield
  {journal} {\bibinfo  {journal} {New Journal of Physics}\ }\textbf {\bibinfo
  {volume} {11}},\ \bibinfo {pages} {045011} (\bibinfo {year}
  {2009})}\BibitemShut {NoStop}%
\bibitem [{\citenamefont {Quan}\ \emph {et~al.}(2016)\citenamefont {Quan},
  \citenamefont {Zhai}, \citenamefont {Wang}, \citenamefont {Hou},
  \citenamefont {Wang}, \citenamefont {Xiang}, \citenamefont {Liu},
  \citenamefont {Zhang},\ and\ \citenamefont {Dong}}]{quan:16}%
  \BibitemOpen
  \bibfield  {author} {\bibinfo {author} {\bibfnamefont {R.}~\bibnamefont
  {Quan}}, \bibinfo {author} {\bibfnamefont {Y.}~\bibnamefont {Zhai}}, \bibinfo
  {author} {\bibfnamefont {M.}~\bibnamefont {Wang}}, \bibinfo {author}
  {\bibfnamefont {F.}~\bibnamefont {Hou}}, \bibinfo {author} {\bibfnamefont
  {S.}~\bibnamefont {Wang}}, \bibinfo {author} {\bibfnamefont {X.}~\bibnamefont
  {Xiang}}, \bibinfo {author} {\bibfnamefont {T.}~\bibnamefont {Liu}}, \bibinfo
  {author} {\bibfnamefont {S.}~\bibnamefont {Zhang}}, \ and\ \bibinfo {author}
  {\bibfnamefont {R.}~\bibnamefont {Dong}},\ }\href {\doibase
  https://doi.org/10.1038/srep30453} {\bibfield  {journal} {\bibinfo  {journal}
  {Scientific reports}\ }\textbf {\bibinfo {volume} {6}},\ \bibinfo {pages}
  {30453} (\bibinfo {year} {2016})}\BibitemShut {NoStop}%
\bibitem [{\citenamefont {Kwiat}\ \emph {et~al.}(1995)\citenamefont {Kwiat},
  \citenamefont {Mattle}, \citenamefont {Weinfurter}, \citenamefont
  {Zeilinger}, \citenamefont {Sergienko},\ and\ \citenamefont
  {Shih}}]{Kwiat:1995ub}%
  \BibitemOpen
  \bibfield  {author} {\bibinfo {author} {\bibfnamefont {P.~G.}\ \bibnamefont
  {Kwiat}}, \bibinfo {author} {\bibfnamefont {K.}~\bibnamefont {Mattle}},
  \bibinfo {author} {\bibfnamefont {H.}~\bibnamefont {Weinfurter}}, \bibinfo
  {author} {\bibfnamefont {A.}~\bibnamefont {Zeilinger}}, \bibinfo {author}
  {\bibfnamefont {A.~V.}\ \bibnamefont {Sergienko}}, \ and\ \bibinfo {author}
  {\bibfnamefont {Y.}~\bibnamefont {Shih}},\ }\href
  {http://link.aps.org/doi/10.1103/PhysRevLett.75.4337} {\bibfield  {journal}
  {\bibinfo  {journal} {Phys. Rev. Lett.}\ }\textbf {\bibinfo {volume} {75}},\
  \bibinfo {pages} {4337} (\bibinfo {year} {1995})}\BibitemShut {NoStop}%
\bibitem [{\citenamefont {Scarani}\ \emph {et~al.}(2009)\citenamefont
  {Scarani}, \citenamefont {Bechmann-Pasquinucci}, \citenamefont {Cerf},
  \citenamefont {Du\ifmmode~\check{s}\else \v{s}\fi{}ek}, \citenamefont
  {L\"utkenhaus},\ and\ \citenamefont {Peev}}]{scarani:09}%
  \BibitemOpen
  \bibfield  {author} {\bibinfo {author} {\bibfnamefont {V.}~\bibnamefont
  {Scarani}}, \bibinfo {author} {\bibfnamefont {H.}~\bibnamefont
  {Bechmann-Pasquinucci}}, \bibinfo {author} {\bibfnamefont {N.~J.}\
  \bibnamefont {Cerf}}, \bibinfo {author} {\bibfnamefont {M.}~\bibnamefont
  {Du\ifmmode~\check{s}\else \v{s}\fi{}ek}}, \bibinfo {author} {\bibfnamefont
  {N.}~\bibnamefont {L\"utkenhaus}}, \ and\ \bibinfo {author} {\bibfnamefont
  {M.}~\bibnamefont {Peev}},\ }\href {\doibase 10.1103/RevModPhys.81.1301}
  {\bibfield  {journal} {\bibinfo  {journal} {Rev. Mod. Phys.}\ }\textbf
  {\bibinfo {volume} {81}},\ \bibinfo {pages} {1301} (\bibinfo {year}
  {2009})}\BibitemShut {NoStop}%
\bibitem [{\citenamefont {Wertheim}\ \emph {et~al.}(1974)\citenamefont
  {Wertheim}, \citenamefont {Butler}, \citenamefont {West},\ and\ \citenamefont
  {Buchanan}}]{Wertheim:1974}%
  \BibitemOpen
  \bibfield  {author} {\bibinfo {author} {\bibfnamefont {G.~K.}\ \bibnamefont
  {Wertheim}}, \bibinfo {author} {\bibfnamefont {M.~A.}\ \bibnamefont
  {Butler}}, \bibinfo {author} {\bibfnamefont {K.~W.}\ \bibnamefont {West}}, \
  and\ \bibinfo {author} {\bibfnamefont {D.~N.~E.}\ \bibnamefont {Buchanan}},\
  }\href {\doibase 10.1063/1.1686503} {\bibfield  {journal} {\bibinfo
  {journal} {Review of Scientific Instruments}\ }\textbf {\bibinfo {volume}
  {45}},\ \bibinfo {pages} {1369} (\bibinfo {year} {1974})}\BibitemShut
  {NoStop}%
\bibitem [{\citenamefont {Rider}(1960)}]{Rider:1960}%
  \BibitemOpen
  \bibfield  {author} {\bibinfo {author} {\bibfnamefont {P.~R.}\ \bibnamefont
  {Rider}},\ }\href {\doibase 10.1080/01621459.1960.10482056} {\bibfield
  {journal} {\bibinfo  {journal} {Journal of the American Statistical
  Association}\ }\textbf {\bibinfo {volume} {55}},\ \bibinfo {pages} {148}
  (\bibinfo {year} {1960})}\BibitemShut {NoStop}%
\bibitem [{\citenamefont {Xu}\ and\ \citenamefont {Xu}(2016)}]{Xu:2016ji}%
  \BibitemOpen
  \bibfield  {author} {\bibinfo {author} {\bibfnamefont {G.}~\bibnamefont
  {Xu}}\ and\ \bibinfo {author} {\bibfnamefont {Y.}~\bibnamefont {Xu}},\ }\href
  {\doibase 10.1007/978-3-662-50367-6} {\emph {\bibinfo {title} {GPS}}},\
  Theory, Algorithms and Applications\ (\bibinfo  {publisher} {Springer Berlin
  Heidelberg},\ \bibinfo {address} {Berlin, Heidelberg},\ \bibinfo {year}
  {2016})\BibitemShut {NoStop}%
\bibitem [{\citenamefont {Ekert}(1991)}]{ekert:92}%
  \BibitemOpen
  \bibfield  {author} {\bibinfo {author} {\bibfnamefont {A.~K.}\ \bibnamefont
  {Ekert}},\ }\href {\doibase 10.1103/PhysRevLett.67.661} {\bibfield  {journal}
  {\bibinfo  {journal} {Phys. Rev. Lett.}\ }\textbf {\bibinfo {volume} {67}},\
  \bibinfo {pages} {661} (\bibinfo {year} {1991})}\BibitemShut {NoStop}%
\bibitem [{\citenamefont {Troupe}\ and\ \citenamefont
  {Lamas-Linares}(2018)}]{Troupe:2018}%
  \BibitemOpen
  \bibfield  {author} {\bibinfo {author} {\bibfnamefont {J.~E.}\ \bibnamefont
  {Troupe}}\ and\ \bibinfo {author} {\bibfnamefont {A.}~\bibnamefont
  {Lamas-Linares}},\ }\href {https://arxiv.org/abs/1808.09019} {\bibfield
  {journal} {\bibinfo  {journal} {arXiv preprint arXiv:1808.09019}\ } (\bibinfo
  {year} {2018})}\BibitemShut {NoStop}%
\end{thebibliography}
%

\end{document}